\documentstyle[aps]{revtex}
\narrowtext
\draft

\begin{document}
\title{{\it Ab initio} calculations of quasiparticle band structure in correlated
systems: LDA++ approach}
\author{A.I. Lichtenstein$^1$ and M.I. Katsnelson$^2$}
\address{$^1$Forschungszentrum J\"ulich, Germany \\
$^2$ Institute of Metal Physics,\\
Ekaterinburg 620219, Russia}
\maketitle

\begin{abstract}
We discuss a general approach to a realistic theory of the electronic
structure in materials containing correlated d- or f- electrons. The main
feature of this approach is the taking into account the energy dependence of
the electron self-energy with the momentum dependence being neglected (local
approximation). It allows us to consider such correlation effects as the
``non-Fermi-step'' form of the distribution function, the enhancement of the
effective mass including ``Kondo resonances'', the appearance of the
satellites in the electron spectra etc. To specify the form of the
self-energy, it is useful to distinguish (according to the ratio of the
on-site Coulomb energy U to the bandwidth W) three regimes - strong,
moderate and weak correlations. In the case of strong interactions (U/W $>$
1 - ''rare-earth system``) the Hubbard-I approach is the most suitable.
Starting from an exact atomic Green function with the constrained density
matrix n$_{mm^{\prime }}$ the band structure problem is formulated as the
functional problem on n$_{mm^{\prime }}$ for f-electrons and the standard
LDA-functional for delocalized electrons. In the case of moderate
correlations (U/W$\sim $1 -''metal -insulator regime, Kondo systems``) we
start from the d$=\infty $ dynamical mean field iterative perturbation
scheme (IPS) of G. Kotliar et. al. and also make use of our multiband atomic
Green function for constrained n$_{mm^{\prime }}$. Finally for the weak
interactions (U/W $<$ 1 -''transition metals``) the self-consistent
diagrammatic fluctuation-exchange (FLEX)-approach of N. Bickers and D.
Scalapino is generalized to the realistic multiband case. We presents
two-band, two-dimensional model calculations for all three regimes. A
realistic calculation in Hubbard-I scheme with the exact solution of the
on-site multielectron problem for f(d)- shells was performed for
mixed-valence 4f compound TmSe, and for the classical Mott insulator NiO.
\end{abstract}


%\widetext

\section{Introduction}

A general accurate description of the electronic structure of materials with
correlated electrons has yet to be developed. Such materials include the
high-T$_c$ and colossal magnetoresistance (CMR) materials, as well as the
mixed-valence and heavy-fermion compounds. All these systems demonstrate
essentially many-particle (correlation) features in their excitation
spectrum and ground-state properties, the usual language of one-electron
band theory being inadequate to describe such features even qualitatively:
e.g., the problem of Mott insulators, the heavy-fermion behavior in some
rare-earth compounds, satellites and ``mid-gap states'' in electron spectra
etc. (see e.g. recent reviews \cite{VKT,PWA-rev,Kotliar-rev,Scalapino-rev}).
Such effects as the metal-insulator transition, Kondo effect, and others,
which helps to understand the basic physics in these strongly correlated
materials, is usually considered in the framework of simplified models such
as the Hubbard model, Anderson model, s-f exchange model and other
correlation models. Nevertheless the complexity of the crystals containing
10-15 different atoms per unit cell, interactions between electronic and
lattice degrees of freedom demands a more detailed investigation of the
energy bands in such systems. The only general first-principles approach
that takes into account in practice specific peculiarities of the electronic
structure in real compounds are those based on the density functional theory
(DFT) \cite{Hohenberg}. The vast majority of practical DFT applications
today are based on the local density approximation (LDA), which treats the
exchange-correlation (XC) part of an effective single-particle DFT potential
as a density dependent XC-potential, taken from the exact quantum
Monte-Carlo (QMC) results for the homogeneous electron gas. There are many
successes but also some failures of LDA approach\cite{Gunnarsson-Jones},
related to the simple fact that in cases where some portion of electronic
structure is better described in terms of atomic-like electronic states, the
homogeneous electron gas approximation is not a good starting point. Another
limitation of LDA theory is that it is only a ground state scheme and the
one-particle band-structure itself has, generally speaking, no proper
meaning. Recently the time-dependent (TD-LDA) approach has been applied for
calculations of excitation energies\cite{Gross,Eguiluz}, but the TD-LDA
effective potential is not known as good as LDA one. On the other hand,
exact QMC calculations for real materials which have tedious first-principle
Hartree-Fock band structure as zero order approximations is still a
challenging problem of solid state theory\cite
{QMC-Urbana,QMC-Jap,QMC-Cambridge}.

In this situation it is useful to have simple and accurate scheme that could
still capture the most important properties of real electronic structure and
at the same time could take into account the most important correlation
effects. One of the first successful approach in this line was the
GW-approximation \cite{Hedin} for quasiparticle spectra in solids with the
self-energy related to a ``bare'' Green function (G) and a screened Coulomb
interaction (W). Self-consistent GW calculations basing on the LDA band
structure gives much better description of a Mott insulator such as NiO than
does pure LDA \cite{Aryasetiawan}. Still, the non-local Coulomb interactions
make such type of calculations really time consuming. For the purpose of
only band-structure investigations one could use a simplified
time-independent GW-scheme or so-called screen-exchange (sX-LDA) approach%
\cite{Rucker,Engel}. The later approach has of the same drawbacks as does
Hartree-Fock approximation and did not suitable for strongly correlated
systems. A different way to incorporate some correlation effect in the
systems with localized d- or f-states was successfully done in so-called
LDA+U method\cite{Anisimov-LDAU}. In this case a simple mean-field
Hubbard-like term is added to LDA functionals for the localized state and
care must be taken for correction of the LDA double-counting \cite
{Solovyev-DC}. This approach can also be viewed as a density functional
theory, since the U terms that depend on occupation number for localized
electrons is a functional of the total density. So one just uses the
LDA-functional for delocalized electrons and improved LDA+U functional for
localized atomic-like states. This approach produces a more reliable
description of the electronic and crystal structure of correlated materials
with charge, spin, and orbital ordering than does the LSDA scheme\cite
{Anisimov-rev}. But LDA+U scheme, as well as the approach based on the so called
self-interaction corrections (SIC) \cite{SIC1}, has one intrinsic
shortcoming related with a mean-field approximation.
It is well known (see e.g. \cite{AGD,Mahan}) that the most
interesting correlation effects in quasiparticle spectra such as the mass
enhancement, damping, the difference of the distribution function from the
``Fermi step'' are connected with the energy dependence of the
self-energy $\Sigma (\omega )$, so one needs to generalize LDA+U approach to
include dynamical effects. 
Such scheme we would like to call ``LDA++''\cite{APS}.
One can mentions few successful attempts in this directions:
quasiparticle (QP) band structure calculation of Fe, Co and Ni\cite
{Kleinman,Treglia,Steiner-3d} as well as heavy fermion system\cite
{Steiner-4f,Sandalov} using simplest second order local approximation for
the self-energy; QP-band structure of NiO\cite{Mangi,Igarashi} using 3-body
Faddeev approximation; RPA-like approach for HTSC\cite{Lopez} and
non-crossing approximations (NCA) for Kondo systems\cite{Han}. At the same
time, a criterion for the applicability of specific approximations used in
these works was not clear.

In this paper we propose a general scheme for LDA++ band structure
calculations for real materials with the different strength of electronic
correlations. It is not very efficiently from the computational point of
view (as well as not very reasonable from the purely theoretical one) to use
the only LDA++ scheme for materials with a different electron-electron
interactions. In accordance with the ratio of average on-site Coulomb
parameter U to the relevant valence band width W, it is useful to
distinguish three regimes of weak, moderate and strong correlations.

The simplest case of weak correlations (U/W$<$1 - ``transition metals'') we
could use the self-consistent diagrammatic approach. The most convenient way
is the conserving fluctuation exchange (FLEX) approximation \cite
{Bickers-FLEX} and we will use the multiband generalizations for LDA++ weak
correlation scheme. The characteristic feature of this renormalized band
regime that no additional states appears in the electronic structure due to
interactions or, more exactly, there is one-to-one correspondence of
quasiparticle states with and without interaction. Roughly speaking, the
shape of energy bands may be changed but there are no band splitting, or
presence of additional bands. The most interesting physical phenomena in
this case are the renormalization of the effective masses, flattening of Van
Hove singularities etc.

In the case of very strong interactions (U/W$>$1 - ``rare-earth system'') we
will start with exact atomic Greens function for f-states and use the
Hubbard-I approximation (HIA) \cite{Hub1} to analyze the spectrum of
f-systems. This approach also may be applied to such d-systems as Mott
insulators with very narrow d-bands. For this situation the electronic
structure of solids will combine the many-body structure of f(d)-ions
and broad bands from delocalized electrons. In this case such phenomena as
multiplet structure, satellites in photoelectron spectra, the narrowing of the
electron bands depending on the magnetic ordering etc are the subjects of
main interests.

In the most difficult case of strongly correlated physics (U/W $\sim $1 -
``metal-insulator transition'' regime or ``Kondo systems'') we will use the
interpolation scheme based on dynamical mean field theory (DMFT)\cite
{Kotliar-rev}. In this situation we will have a ``three-peak structure'' from a
single correlated band, consisting of upper and lower Hubbard bands and
Kondo resonance near the Fermi energy. Such scheme is the most accurate, 
but also the most time consuming, and it is difficult to make a
self-consistent calculations for a large system. The other point of view of
the different LDA++ schemes could be related to the different energy scales
for the spectrum of correlated materials: if one is interested in the large
energy scale that HIA-approximation is sufficient for spectroscopic
purposes. If we like to describe low-energy scale of a system like doped
Mott insulators or mixed valence systems then DMFT approach is the most
adequate.

The common feature of all LDA++ methods is the matrix form of self-energy
since electron-electron correlations can not be diagonalized neither in band
index $n$ nor in orbital indices $lm$. This peculiarity of multi-band
Hubbard interactions are normally ignored and only few examples of matrix
self-energy exist for transition metal with LDA second-order perturbation
scheme \cite{Kleinman,Steiner-3d}, transition-metal oxides within 3-body
Faddeev approximation\cite{Igarashi} and for two-band Hubbard model for
investigation of orbital and magnetic instabilities \cite{Buda-RPA}.

This paper is organized as follows. In Sec. II. we will give a general
description of different correlations scheme to the dynamical mean field
band structure calculations. The simple two-bands model (Sec.III) will
illustrate in practice all LDA++ methods. In the Sec.IV we will give an
example of first-principle calculations of mixed-valence system TmSe and
classical Mott insulator NiO within Hubbard-1 approximation to LDA++.
Finally we summarize our results in the sec V.

\section{LDA++ methods}

The ``Kohn-Sham'' energies of one-particle LDA states
cannot be considered as the quasiparticle energies in the sense of
many-particle theory (see e.g. \cite{AGD,Mahan}).
In the LDA++ approach they considered  
only as the ``bare'' energies and are supposed to be renormalized by the
correlation effects. Of course they contain already some part of the
correlation effects but only those which may be considered in the local
density approximation. The most important ``rest'' in strongly correlated
system is the correlations of the Hubbard type \cite{Hub1} due to the
intra-site Coulomb repulsion. Therefore our starting point is the same as in
the LDA+U approach. We proceed with the Hamiltonian

\begin{equation}
H=\sum_{ij\sigma \{m\}}^{}t_{m_1m_2}^{ij}c_{im_1\sigma }^{+}c_{jm_2\sigma }+%
\frac 12\sum_{i\{\sigma m\}}U_{m_1m_2m_1^{^{\prime }}m_2^{^{\prime
}}}^ic_{im_1\sigma }^{+}c_{im_2\sigma ^{^{\prime }}}^{+}c_{im_2^{^{\prime
}}\sigma ^{^{\prime }}}c_{im_1^{^{\prime }}\sigma }  \label{H-mb}
\end{equation}
where the ($i,j$) represents different crystal sites, \{$m$\} label
different orbitals and the \{$\sigma $\} are spin indices. Coulomb matrix
elements are defined in the usual way:

\begin{equation}
U_{m_1m_2m_1^{^{\prime }}m_2^{^{\prime }}}=\int \int d{\bf r}d{\bf r}%
^{^{\prime }}\psi _{m_1}^{*}({\bf r})\psi _{m_2}^{*}({\bf r}^{\prime
})V_{ee}({\bf r}-{\bf r}^{\prime })\psi _{m_1^{^{\prime }}}({\bf r})\psi
_{m_2^{^{\prime }}}({\bf r}^{\prime })  \label{U-mat}
\end{equation}
here $V_{ee}({\bf r}-{\bf r}^{\prime })$ is the screened Coulomb
interactions and $\psi _m({\bf r})$ are localized on-site basis functions
(the site index being suppressed).

In this case all the orbitals are assumed to belong to the ``correlated''
set, while in real materials like high-T$_c$ compounds (e.g., YBa$_2$Cu$_3$O$%
_{7\text{ }}$) we may define as a first approximation only the 3d orbitals
as correlated ones. Therefore it is more reasonable to rewrite Eq.\ref{H-mb}
in the form of LDA+U Hamiltonian:

\begin{equation}
H=H_{dc}^{LDA}+\frac 12\sum_{i\{\sigma m\}}U_{m_1m_2m_1^{^{\prime
}}m_2^{^{\prime }}}^ic_{im_1\sigma }^{+}c_{im_2\sigma ^{^{\prime
}}}^{+}c_{im_2^{^{\prime }}\sigma ^{^{\prime }}}c_{im_1^{^{\prime }}\sigma}
\label{LDAU}
\end{equation}

Note that index $i$ in the second sum of Eq.\ref{LDAU} running only for
``correlated'' sites, and orbital indices $\left\{ m\right\} $ only for
''correlated'' states (e.g. 3d or 4f) while the first LDA-term:

\[
H_{dc}^{LDA}=\sum_{ij\sigma \{m\}}h_{m_1m_2\sigma }^{ij}c_{im_1\sigma
}^{+}c_{jm_2\sigma }-E_{dc} 
\]
contains all sites and orbitals in the unit cell. Here $h_{m_1m_2\sigma
}^{ij}$ are the ``one-particle'' Hamiltonian parameters in the
(spin-polarized) LDA, $E_{dc}$ is the double counting correction for average
Coulomb interactions in L(S)DA \cite{Solovyev-DC}:

\[
E_{dc}=\frac 12\overline{U}n_d(1-n_d)-\frac 12\overline{J}[n_{d\uparrow
}(1-n_{d\uparrow })+n_{d\downarrow }(1-n_{d\downarrow }) 
\]
with $\overline{U}$ and $\overline{J}$ being average Coulomb and
exchange interactions and $n_d=n_{d\uparrow }+n_{d\downarrow }$ is the total
number of correlated d(f)-electrons.

One non-trivial problem is to find an efficient way to compute the one
electron Hamiltonian $h^{LDA}$ in a minimal orthogonal basis set. Orthogonality
of the basis functions is required for the use of second quantization form
of the effective Hamiltonian (Eq.\ref{H-mb}). Since many-body ($U$) part of
the problem is an order of magnitude more time consuming than LDA one, we
need to use a minimal basis set and integrated out all high-energy degrees
of freedom (out of the $\pm U$ range). One of the best LDA-methods for such
scheme is the Linear Muffin-Tin Orbital (LMTO) method \cite{OKA-exact},
which could give the orthogonal down-folded one-electron tight-binding
Hamiltonian.
In this case LDA calculations corrected for double
counting, produce the first-principle hopping $t_{ij}$ in many body
Hamiltonian.

Now we can describe methods to efficient calculations of quasi-particle (QP)
spectra for the LDA+U Hamiltonian. In this sense our approach is not any
more density functional theory and one could benefit from the possibility to
use the information on QP-band structure comparing with different
``excitation'' experiments.

\subsection{Multi-band FLEX}

In this section we generalize the FLEX equations\cite
{Bickers-FLEX} for the purpose of multiband LDA++ scheme. We will not take
into account a momentum ${\bf q}$-dependence of the self-energy, although in
the FLEX approximation it is straightforward to include it in all the
following formulas. The numerical computation of the $({\bf q},\omega )$%
-dependent self-energy is time-consuming in multiband case \cite
{Licht-QP,Putz-3bFLEX}. To unify the approximations for all our LDA++
schemes we will not include explicitly the ${\bf q}$-dependence in the FLEX
formalism.

First of all one needs to symmetrize the bare vertex matrix U over different
fluctuation channels: particle-hole (density - U$^d$ and magnetic - U$^m$)
and particle-particle (singlet - U$^s$ and triplet - U$^t$)
vertex matrix:

\begin{eqnarray*}
U_{m_1m_1^{^{\prime }}m_2m_2^{^{\prime }}}^d&=&2U_{m_1m_2m_1^{^{\prime
}}m_2^{^{\prime }}}-U_{m_1m_2m_2^{^{\prime }}m_1^{^{\prime }}} \\
U_{m_1m_1^{^{\prime }}m_2m_2^{^{\prime }}}^m&=&-U_{m_1m_2m_2^{^{\prime
}}m_1^{^{\prime }}} \\
U_{m_1m_1^{^{\prime }}m_2m_2^{^{\prime }}}^s&=&\frac 12(U_{m_1m_1^{^{\prime
}}m_2m_2^{^{\prime }}}+U_{m_1m_1^{^{\prime }}m_2^{^{\prime }}m_2}) \\
U_{m_1m_1^{^{\prime }}m_2m_2^{^{\prime }}}^t&=&\frac 12(U_{m_1m_1^{^{\prime
}}m_2^{^{\prime }}m_2}-U_{m_1m_1^{^{\prime }}m_2m_2^{^{\prime }}})
\end{eqnarray*}

The one-electron Green-function is defined through the following equation:

\[
G_{mm^{^{\prime }}\sigma }^{-1}(i\omega _n )=(i\omega _n +\mu )\delta
_{mm^{^{\prime }}}-h_{mm^{^{\prime }}\sigma }-\Sigma _{mm^{^{\prime }}\sigma
}^{HF}-\Sigma _{mm^{^{\prime }}\sigma }(i\omega _n ) 
\]
here $\mu$ is chemical potential, 
 $\omega _n=(2n+1)/\beta $ are Matsubara frequencies and $\beta =1/k_BT$
is inverse temperature. The frequency independent Hartree-Fock part is

\begin{equation}
\Sigma _{mm^{^{\prime }}\sigma }^{HF}=\sum_{m_1m_2}(U_{mm_1m^{^{\prime
}}m_2}\sum_{\sigma ^{^{\prime }}}n_{m_1m_2}^{\sigma ^{^{\prime
}}}-U_{mm_1m_2m^{^{\prime }}}n_{m_1m_2}^\sigma )  \label{Sig-HF}
\end{equation}
and corresponds to the rotationally invariant LDA+U method \cite
{Licht-RILDAU}.

It is useful to write the multi-band FLEX equations using matrix-vector
notation for different Coulomb matrix vertices and vector Green function. We
will use a combine index: $\alpha =\{m,m^{^{\prime }}\}$ and defined the
vector Green function as well as matrix interactions in the following way:

\[
{\bf G}\equiv \{G_\alpha \}, \,\,\, \widehat{U}=\{U_{\alpha \alpha
^{^{\prime }}}\} 
\]
For simplicity we first write equations for non-polarized spin states and
omit the spin indices. In this case the Hartree-Fock approximation Eq.\ref
{Sig-HF} can be rewritten in the form of a matrix-vector product only with
``density'' Coulomb interaction:

\[
{\bf \Sigma }^{HF}=\widehat{U}^d*{\bf n} 
\]
where the occupation matrix is defined as:

\[
n_\alpha \equiv n_{mm_{^{\prime }}}^\sigma =<c_{m\sigma }^{+}c_{m^{^{\prime
}}\sigma }>=\frac 1\beta \sum_{\omega _n}G_{m^{\prime }m}(i\omega _n)+\frac 1%
2\delta _{mm^{^{\prime }}} 
\]

Using the single-site Hubbard interactions one obtains a ``local form''
of FLEX equations in the frequency ($\omega $) - time ($\tau $) space. It is
very efficient to use fast-Fourier transforms with periodic boundary
condition\cite{Serene-FFT}. Time-frequency spaces are connected by

\begin{eqnarray*}
{\bf G}(i\omega _n) &=&\int_0^\beta e^{i\omega _n\tau }{\bf G}(\tau )d\tau \\
{\bf G}(\tau ) &=&\frac 1\beta \sum_{\omega _n}e^{-i\omega _n\tau }{\bf G}%
(i\omega _n)
\end{eqnarray*}

We will try to keep this dual ($\omega -\tau )$ notation to stress the
numerical implementation of this LDA++ scheme. We write the approximation for
self-energy in the ``$GW$''-like form :

\begin{equation}
{\bf \Sigma }(\tau )=\widehat{W}(\tau )*{\bf G}(\tau )  \label{GW}
\end{equation}
where symmetrized fluctuation $W(i\omega )$-potential defined as:

\[
W_{m_1m_2m_1^{^{\prime }}m_2^{^{\prime }}}=V_{m_1m_1^{^{\prime
}},m_2^{^{\prime }}m_2} 
\]
and total fluctuation potential consists of the second-order term,
as well as particle-hole and particle-particle contributions:

\[
\widehat{V}(i\omega )=\widehat{V}_2(i\omega )+\widehat{V}_{ph}(i\omega )-%
\widehat{V}_{pp}(-i\omega ) 
\]

All these contributions can be expressed in terms of bare $(D_0, M_0, S_0, T_0)$
and renormalized $(D, M, S, T)$ channel propagators. The second order
potential for the non-magnetic case is

\begin{equation}
\widehat{V}_2(i\omega )=\widehat{U}*\widehat{D}_0(i\omega )*\widehat{U}^d
\label{V2}
\end{equation}
while the particle-hole potential is expressed through the density and
magnetic fluctuations:

\[
\widehat{V}_{ph}(i\omega )=\frac 12\widehat{U}^d*[\widehat{D}(i\omega )-%
\widehat{D}_0(i\omega )]*\widehat{U}^d+\frac 32\widehat{U}^m*[\widehat{M}%
(i\omega )-\widehat{M}_0(i\omega )]*\widehat{U}^m 
\]
Finally the particle-particle contribution to fluctuation-exchange potential
is:

\[
\widehat{V}_{pp}(i\omega )=\widehat{U}^s*[\widehat{S}(i\omega )-\widehat{S}%
_0(i\omega )]*\widehat{U}^s+3\widehat{U}^t*[\widehat{T}(i\omega )-\widehat{T}%
_0(i\omega )]*\widehat{U}^t 
\]
If one defines the particle-hole ($\chi $) and particle-particle ($\pi $)
''empty loop'' susceptibilities:

\begin{equation}
\chi _{m_1m_2m_3m_4}(\tau )=-G_{m_4m_1}(-\tau )*G_{m_2m_3}(\tau )
\label{susc}
\end{equation}

\[
\pi _{m_1m_2m_3m_4}(\tau )=G_{m_1m_4}(\tau )*G_{m_2m_3}(\tau ) 
\]
we can write with this notations for susceptibilities the bare channel
propagator matrices in the following form for the density and magnetic part:

\[
\widehat{D}_0=\widehat{M}_0=\widehat{\chi } 
\]
and for singlet and triplet bare propagators:

\[
S_{m_1m_2m_3m_4}^0=\frac 12(\pi _{m_1m_2m_3m_4}+\pi _{m_1m_2m_4m_3}) 
\]

\[
T_{m_1m_2m_3m_4}^0=\frac 12(\pi _{m_1m_2m_3m_4}-\pi _{m_1m_2m_4m_3}) 
\]
The total channel propagators (R$_\lambda $ where $\lambda $=\{d, m, s, t\})
have to be found from the RPA-like matrix inversion:

\[
\widehat{R}_\lambda (i\omega )=[\widehat{1}+\widehat{R}_\lambda ^0(i\omega
)*U^\lambda ]^{-1}*\widehat{R}_\lambda ^0(i\omega ) 
\]

The derivation of the complete expression for the FLEX self energy for
spin-polarized case with taking into account all the channels are rather
cumbersome. Since we will not use here these complicated expressions, they
will be discussed in details elsewhere\cite{SPFLEX}.

\subsection{Hubbard-I approximation}

Historically Hubbard-I approximation \cite{Hub1} was the first and the
simplest approximations for a strongly correlated one-band model. It has,
however, many inconsistencies (see e.g. discussion in \cite{Gebhard}). For
example it is not conserving (the self energy cannot be represented as a
functional derivative of the generating functional with respect to the Green
function) and therefore does not obey the Luttinger theorem and other
``exact'' Fermi-liquid properties. For the half-filled nondegenerate Hubbard
model it always gives a gap in the energy spectrum, even for small U. This
means that HIA is completely inapplicable for small and medium interactions. But at
the same time it gives a correct picture of the electron spectrum in the
narrow-band limit. Therefore it seems to be very useful in 4f systems with a
very strong degree of localization of the electron states. Applying to some
real systems in the framework of LDA++ approach, HIA-scheme could give (as
it will be shown below) an effective and non-trivial descriptions of
many-body multiplet effects.

To introduce Hubbard-I type approximation in degenerate case it is
convenient to exploit the so called atomic representation and Hubbard $X$%
-operators (see \cite{Hub4,VKT,II})

\[
X_i^{\mu \nu }=\left| i\mu \right\rangle \left\langle i\nu \right| 
\]
where $\mu ,\nu $ are multielectron states of the site $i$ as a whole
(configuration and multiplet indices). In terms of $X$-operators the atomic
Hamiltonian has very simple form

\[
H^{at}=\sum_\mu E_\mu X^{\mu \mu } 
\]

On the other hand, the inter-site transfer Hamiltonian, which has very
simple (bilinear) structure in terms of the operators $c_{m^{^{\prime
}}\sigma }^{+}$, $c_{m\sigma }$, also can be expressed in terms of $X$%
-operators by the relations $c_{m\sigma }=\sum_{\mu \nu }\left\langle \mu
\left| c_{m\sigma }\right| \nu \right\rangle X^{\nu \mu }$ and similarly for 
$c_{m^{^{\prime }}\sigma }^{+}.$ In the limit of a very strong interaction
it is convenient to calculate the Green function via $X$-operators (using
the decoupling procedure \cite{II} or special diagram technique for $X$%
-operators \cite{Izumov}) and then transform to electron operators. HIA
corresponds to the following expression \cite{Hub1,II}:

\[
G^{-1}\left( i\omega \right) =\left[ G^{at}\left( i\omega \right) \right]
^{-1}-\widehat{t} 
\]
where $\widehat{t}$ is the matrix of transfer integrals. In the limit of
very small $\widehat{t}$ this expression describes the arising of separate
bands from each intraatomic transition with the change of the electron
number from unity. It is the picture that seems reasonable for e.g.
rare-earth materials with very narrow 4f-band. The bands always appear to be
narrowed. Indeed, if in the vicinity of the pole $i\omega =\varepsilon _0$
the atomic Green function can be represented as

\[
G^{at}\left( i\omega \right) =\frac{Z_0}{i\omega -\varepsilon _0} 
\]
the effective transfer Hamiltonian for this ``Hubbard band'' will be $Z_0%
\widehat{t}$ instead of $\widehat{t}$.

In terms of LDA++ multiband approach HIA for Green function has the following form

\begin{equation}
G_{im,jm^{^{\prime }},\sigma }^{-1}(i\omega ) =\left[ (i\omega+\mu)\delta
_{mm^{^{\prime }}}-\Sigma _{mm^{^{\prime }}\sigma }^{at}(i\omega )\right]
\delta _{ij}-h_{mm^{^{\prime }}\sigma }^{ij}  \label{Hub-1}
\end{equation}
To obtain this Green function, we need to solve by an exact diagonalization
(ED) technique the atomic many-electron problem:

\[
H^{at}|\upsilon >=E_\nu ^{at}|\upsilon > 
\]
with the effective atomic Hamiltonian for d- or f-states:

\begin{equation}
H^{at}=\sum_{mm^{^{\prime }}\\\sigma }\varepsilon _{mm^{^{\prime
}}}c_{m\sigma }^{+}c_{m^{^{\prime }}\sigma }+\frac 12\sum_{\{\sigma
m\}}U_{m_1m_2m_1^{^{\prime }}m_2^{^{\prime }}}c_{im_1\sigma
}^{+}c_{im_2\sigma ^{^{\prime }}}^{+}c_{im_2^{^{\prime }}\sigma ^{^{\prime
}}}c_{im_1^{^{\prime }}\sigma }  \label{H-at}
\end{equation}
Here $\varepsilon _{mm^{^{\prime }}\text{ }}$is the matrix of atomic
energies which in principle can include non-diagonal terms. The latter is
naturally come from LMTO-TB effective Hamiltonian which has a diagonal part
of $h_{mm^{^{\prime }}\sigma }$ as a result of transformation to an
orthogonal basis set\cite{OKA-exact}. Diagonalization of atomic Hamiltonian
Eq.\ref{H-at} is not a big problem for a standard workstation, since it is
equivalent to 5- and 7-site Hubbard model in ED-scheme\cite{Kotliar-rev} for
d- and f-states.

Using eigenfunctions and eigenvectors of Hamiltonian (Eq.\ref{H-at}, the exact
atomic Green-function can be found by the standard definition\cite{AGD}:

\begin{equation}
G_{mm^{^{\prime }}\sigma }^{at}(i\omega )=\frac 1Z\sum_{\mu \nu }\frac{%
\left\langle \mu \left| c_{m\sigma }\right| \nu \right\rangle \left\langle
\nu \left| c_{m^{^{\prime }}\sigma }^{+}\right| \mu \right\rangle }{i\omega
+E_\mu -E_\nu }(e^{-\beta E_\mu }+e^{-\beta E_\nu })  \label{Gat}
\end{equation}

where $Z=\sum_\nu e^{-\beta E_\nu }$.

Finally $\Sigma ^{at}$ which is needed for HIA-approximation is found from
the following expression:

\begin{eqnarray}
\Sigma _{mm^{^{\prime}}\sigma }^{at}(i\omega ) =i\omega \delta
_{mm^{^{\prime }}}-\varepsilon _{mm^{^{\prime }}}
-(G^{at})_{mm^{^{\prime }}\sigma }^{-1}(i\omega )  \label{Self-at} 
\end{eqnarray}

Now the HIA-approach to LDA++ may be formulated as a functional for atomic
density matrix: $n_{mm^{^{\prime }}}$ with a constraint (for $\varepsilon
_{mm^{^{\prime }}}$)

\begin{equation}
n_{mm^{^{\prime }}}=\frac 1\beta \sum_\omega G_{mm^{^{\prime
}}}^{at}(i\omega )+\frac 12\delta _{mm^{^{\prime }}}  \label{N-at}
\end{equation}
(see \cite{AGD}) having the same $n_{mm^{^{\prime }}}$ density matrix for d-
or f-electrons as in the crystal as for the corresponding site and orbital
element of the Green function, Eq.\ref{Hub-1}.

\subsection{DMFT-multiband scheme.}

A great success of dynamical mean field (d=$\infty $) approach to the theory
of correlated systems \cite{Kotliar-rev} shows that probably this scheme can
be the most accurate for the calculations of the self energy from local
description of electron fluctuations, at least in the vicinity of the metal-insulator
transition. We will use this scheme for real crystals as a best local
approximation. The DMFT scheme is based on the ''cavity method'' or the
solution of the effective impurity problem, which corresponds to subtraction
of the local self energy only on the one atom in question. In appendix A we
show the equivalence of the cavity and impurity methods for matrix multiband
Hamiltonians. It was realized recently that the success of DMFT in the
one-band half-filled Hubbard model with simplest second-order self-energy is
related to the fact that both small and large U-limits are exact in this
case \cite{Kotliar-rev}. This is not true for non-integer filling or for the
multiband case. The elegant iterative perturbation scheme (IPS) for
non-integer one-band Hubbard model was proposed recently \cite{Kajuter-IPT}
and gives almost perfect agreement with ED- and QMC-results. For the case of
multi-band with non-integer occupations the problem is much more severe and
the existing IPS-generalization \cite{Kajuter-mband} does not produce good
results for large doping. Here we use the main idea of the original
IPS-method\cite{Kajuter-IPT} and propose another version of multiband DMFT
which is based more on the numerical solutions of corresponding atomic
problem than the approximate analytical one used in \cite{Kajuter-mband}.

The impurity problem for the ``bath`` Green function reads:

\[
\lbrack {\cal G}_0(i\omega )]_{mm^{^{\prime }}}^{-1}=[G(i\omega
)]_{mm^{\prime }}^{-1}+(\mu _0-\mu )\delta _{mm^{^{\prime }}}+\Sigma
_{mm^{^{\prime }}}(i\omega ) 
\]
where the local Green function is defined through Brillouin zone sum:

\[
G_{mm^{^{\prime }}}(i\omega )=\frac 1{N_k}\sum_{{\bf k}}G_{mm^{^{\prime }}}(%
{\bf k},i\omega ) 
\]
here $N_k$ is the total number of ${\bf k}$-points. Alternatively one may
perform the ${\bf k}$-integration using a complex-tetrahedron scheme \cite
{Lambin-TET,Anis-DMFT}. We introduce here according to \cite{Kajuter-IPT}
the ''local'' impurity chemical potential $\mu _0$ to satisfy the condition

\[
\sum_{{\bf k}}\int\limits_{-\infty }^{+\infty }\frac{d\omega }{2\pi }%
Tr\left[ \widehat{G}({\bf k,}i\omega )\frac{\partial \widehat{\Sigma }%
(i\omega )}{\partial \omega }\right] =0 
\]
which is necessary to provide the Luttinger theorem to be true.

We use the following ansatz for the self energy is in the matrix ($%
m,m^{^{\prime }}$) form:

\begin{equation}
\widehat{\Sigma }(i\omega )=\widehat{\Sigma }^{HF}(i\omega )+\widehat{A}*%
\widehat{\Sigma }^{(2)}(i\omega )*\left[ \widehat{1}-\widehat{B}(i\omega )*%
\widehat{\Sigma }^{(2)}(i\omega )\right] ^{-1}  \label{Sig-DMFT}
\end{equation}
where the second-order self energy $\widehat{\Sigma }^{(2)}(i\omega )$ is
defined in terms of the bath Green function ${\cal G}_0(i\omega )$ in
according with Eq.\ref{GW} with $W=V^{(2)}$ (see also Eq.\ref{V2}):

\[
\widehat{\Sigma }^{(2)}=\widehat{\Sigma }^{(2)}[{\cal G}_0] 
\]

In spirit of the approach \cite{Kajuter-IPT} the $A$-matrix should be
defined to provide the exact high-energy ($\omega \rightarrow \infty )$
limit of $\widehat{\Sigma }(i\omega ).$ The best way to receive such
asymptotic is the using of the equations of motion for the double-time
retarded Green function with the analytical continuation on the Matsubara
frequencies \cite{AGD,Izumov}. One has exact at $\omega \rightarrow \infty $
expression

\[
\Sigma _{mm^{\prime }\sigma }\left( i\omega \right) =\frac 1{i\omega }%
N_{mm^{\prime }\sigma } 
\]
where 
\[
N_{mm^{\prime }\sigma }=\left\langle \left\{ \left[ c_{m\sigma
},H_{int}\right] ,\left[ H_{int},c_{m^{\prime }\sigma }^{+}\right] \right\}
\right\rangle 
\]
here $\left[ ...,...\right] $ and $\left\{ ...,...\right\} $ are the symbols for
commutator and anticommutator, correspondingly, $H_{int}$ is the Hubbard
(interaction) part of the Hamiltonian. Note that in the multiband case the
average $N_{mm^{\prime }\sigma }$ contains the products of four electron
operators and cannot be found exactly. Decoupling of these four-fermion
averages according to the Wick theorem and comparing the result with the
asymptotic of $\widehat{\Sigma }^{(2)}$:

\[
\Sigma _{mm^{^{\prime }}\sigma }^{(2)}(i\omega \rightarrow \infty )=\frac{%
N_{mm^{^{\prime }}\sigma }^0}{i\omega } 
\]
we obtain the following expression

\[
\widehat{A}=\widehat{N}*\left[ \widehat{N}^0\right] ^{-1} 
\]
where $\widehat{N_0}$-matrix defined in the spin-polarized case as:

\begin{eqnarray*}
N_{mm^{^{\prime }}\sigma }^0
&=&\sum_{\{m_i\}}\{U_{mm_3m_1m_4}U_{m_1m_5m^{^{\prime }}m_2}\sum_{\sigma
^{^{\prime }}}n_{m_5m_4\sigma ^{^{\prime }}}^0(\delta
_{m_3m_2}-n_{m_3m_2\sigma ^{^{\prime }}}^0) \\
&&-U_{mm_3m_4m_1}U_{m_1m_5m^{^{\prime }}m_2}n_{m_5m_4\sigma ^{^{\prime
}}}^0(\delta _{m_3m_2}-n_{m_3m_2\sigma }^0)\}
\end{eqnarray*}
and in the non-magnetic case it simplifies to:

\[
N_{mm^{^{\prime }}}^0=\sum_{\{m_i\}}U_{mm_1m_3m_4}^dU_{m_1m_5m^{^{\prime
}}m_2}n_{m_5m_4}^0(\delta _{m_3m_2}-n_{m_3m_2}^0) 
\]
The expression for $\widehat{N}$-matrix differing from that for $\widehat{N}%
^0$ by the replacement of the occupation matrix $n^0\rightarrow n$. Note
that matrix $\widehat{A}$ appears to be non-Hermitian. In the non-degenerate
case this expression appears to be exact (see \cite{Kajuter-IPT}) due to the
identity

\[
\left( c_{m\sigma }^{+}c_{m\sigma }\right) ^2=c_{m\sigma }^{+}c_{m\sigma }. 
\]
It can be quite accurate also in the general multiband case.

Coefficient matrix $\widehat{B}$ is designed to fix the exact atomic limit
of the interaction self energy, Eq.\ref{Sig-DMFT}. There are other problems
with coefficient $B$ in the multiband case \cite{Kajuter-mband}. While in
single band model one can find an analytical expression for the constant  
$B$\cite{Kajuter-IPT}, in the multiband case this parameter should be $\omega $%
-dependent, owing to the frequency dependence of the atomic self-energy, Eq.%
\ref{Self-at}. We decide to find numerically the non-Hermitian matrix $%
B(i\omega )$ from atomic limit of Eq.\ref{Sig-DMFT} using the exact $%
\widehat{\Sigma }_{at}(i\omega )$ with a constraint for the density matrix $%
\widehat{n}$. In this limit $\widehat{\Sigma }^{(2)}(i\omega
) $ in the non-magnetic case has the form:

\[
\widehat{\Sigma }_{mm^{^{\prime }}}^{(2)at}(i\omega
)=\sum_{\{m_i\}}U_{mm_1m_3m_2}^dU_{m_1m_2m^{^{\prime }}m_3}\frac{%
[f_{m_3}(1-f_{m_2}-f_{m_1})+f_{m_1}f_{m_2}]}{i\omega +\mu _0-\varepsilon
_{m_2}+\varepsilon _{m_3}-\varepsilon _{m_1}} 
\]
here $f_{m_i}$ and $\varepsilon _{m_i}$ are diagonal occupation numbers and
energies of $h^{at}.$ In this case we have:

\begin{equation}
\widehat{B}(i\omega )=\left[\widehat{\Sigma }^{(2)at}(i\omega)\right] ^{-1}-
\left[\widehat{\Sigma }^{at}(i\omega)-\widehat{\Sigma }^{HF}\right] ^{-1}* 
\widehat{A}  \label{Bom}
\end{equation}

As a simple example for such scheme we compare on Fig.\ref{toy} DMFT to
exact-diagonalization for the Anderson model of two-site, two-bands with one
correlated site $U=4$, $\varepsilon _f=-4$ and one ``free-site'' with $%
\varepsilon _0=0$ and hybridization between the sites $V=0.25$ \cite
{Kajuter-mband}. For convenient we assume that all parameters for our model
calculations are in ``eV'' energy units. The corresponding Hamiltonian for
Anderson impurity model has the following form:

\[
H_{imp}=\epsilon _f\sum_{m\sigma }f_{m\sigma }^{+}f_{m\sigma
}+V\sum_{m\sigma }(f_{m\sigma }^{+}c_{m\sigma }+c_{m\sigma }^{+}f_{m\sigma
})+U\sum_mf_{m\uparrow }^{+}f_{m\uparrow }f_{m\downarrow }^{+}f_{m\downarrow
} 
\]
It is not a problem to find an exact Green function for this model
(non-symmetrize many-body Hamiltonian has the dimension $256x256$) and compare
with approximate calculations. We see that the agreement between exact
solution and our DMFT results is quite good even for a large filling (n$%
_{tot}>1;$ in this case n$_f=0.76,$ n$_{tot}\approx 2$ ). Also note that the
atomic Green-function in Fig.\ref{toy} for correlated site has the
``three-peak'' structure for this occupations (there are general 8 poles in
Green-functions for two-band case) and not the two-peak structure as in the
one-band model. The use of the {\it numerical} atomic Green function for the $%
B(i\omega )$-matrix calculation is quite important even for qualitative
agreement with exact results for such model at the filling larger than one
electron per site\cite{Kajuter-IPT}

\section{Results for two-band model.}

In this section we compare the three different LDA++ approaches described
above for a two-band system. We used the simplified two-dimensional model
for High-Tc superconductors for $d_{x^2-y^2}\equiv x$ and $d_{z^2}\equiv z$
orbitals \cite{Buda-RPA}. If one can skip $z$-orbital it will be standard
single-band nearest neighbor hoping (``t'')-model. LDA band structure
calculation for High-Tc materials shows the large contribution of Cu $%
d_{z^2} $ orbital to states near the Fermi level\cite{OKA-TB}. Therefore
the situation with two correlated valence bands could be possible in this
materials. Although we knew that for the realistic description of Cu $%
d_{x^2-y^2}$ state in single band model one need to include next-nearest
hoppings (coming from interactions with O2p- and Cu4s-orbitals\cite{OKA-TB})
we used here the simplified tight-binding model for two correlated bands
``x'' and ``z'' within nearest-neighbour hopping approximation. The one electron
Hamiltonian has the following form\cite{Buda-RPA}:

\[
\widehat{h}({\bf k})=\left( 
\begin{array}{cc}
-2t_{xx}(\cos k_x+\cos k_y)+\Delta , & -2t_{xz}(\cos k_x-\cos k_y) \\ 
-2t_{xz}(\cos k_x-\cos k_y), & -2t_{zz}(\cos k_x+\cos k_y)
\end{array}
\right) 
\]
The hopping parameters are related via simple Slater-Koster ratio: $%
t_{xx}=1,\,\,t_{zz}=0.3,\,\,t_{xz}=0.4$. Again we assume that all
TB-parameters are in ``eV'' energy units, while the value $t_{xx}\sim $0.5
eV would be more realistic\cite{OKA-TB}. It is important to take into
account the energy shifting parameter $\Delta $ since Cu $d_{x^2-y^2}$ bands
located higher than Cu $d_{z^2}$-one, so we use $\Delta =4.$ For the Coulomb
energy our parameterization corresponds to the following matrix elements ($%
m_1\neq m_2$): $\,U_{m_1m_1m_1m_1}=U+J,\,\,U_{m_1m_2m_1m_2}=U\,$, $%
\,\,U_{m_1m_2m_2m_1}=J ,$and $\,\,U_{m_1m_2m_1m_1}=\delta J.$ In this case
the symmetrized bare vertices has the following form (the basis function
numbering as: $xx,xz,zx,zz$):

\begin{eqnarray*}
U^d &=&\left( 
\begin{array}{cccc}
U+J & \delta J & \delta J & 2U-J \\ 
\delta J & 0 & 2J-U & \delta J \\ 
\delta J & 2J-U & 0 & \delta J \\ 
2U-J & \delta J & \delta J & U+J
\end{array}
\right) ,\,\,U^m=\left( 
\begin{array}{cccc}
-U-J & -\delta J & -\delta J & -J \\ 
-\delta J & 0 & -U & -\delta J \\ 
-\delta J & -U & 0 & -\delta J \\ 
-J & -\delta J & -\delta J & -U-J
\end{array}
\right) \\
&& \\
U^s &=&\left( 
\begin{array}{cccc}
U+J & \delta J & \delta J & 0 \\ 
\delta J & \frac 12(J+U) & \frac 12(J+U) & \delta J \\ 
\delta J & \frac 12(J+U) & \frac 12(J+U) & \delta J \\ 
0 & \delta J & \delta J & U+J
\end{array}
\right) ,\,\,U^t=\left( 
\begin{array}{cccc}
0 & 0 & 0 & 0 \\ 
0 & \frac 12(J-U) & \frac 12(J-U) & 0 \\ 
0 & \frac 12(J-U) & \frac 12(J-U) & 0 \\ 
0 & 0 & 0 & 0
\end{array}
\right)
\end{eqnarray*}

We investigate this model for different U parameters: $U=2-8$ with the fixed
values of $J=0.5$ and $\delta J=0.1$. The total number of electrons are $%
n_{tot}=1.4$ which approximately corresponds to fully occupied $z$-bands and
almost half-field $x$-band with 10\% of holes. We use the $32\times 32$ mesh
for the summation over the Brillouin zone and $4000-8000$ Matsubara
frequencies with the cutoff energy equals to $20-40$ times the bandwidth. On
the Fig.\ref{dosfl} we show the results of self-consistent FLEX two-band
calculations for $U=2,4,6$ and $8$. Density of state (DOS) was obtained from
the Green function, extrapolating Matsubara frequencies with Pade
approximation \cite{Serene-Pade} to the real axis. Note that the bare
bandwidth for $t_{xz}=0$ is equal to $8t$ in the case of two dimensional
square lattice and corresponds to 8 and 2.4 eV for $x$ and $z$ bands. One
can see the narrowing the DOS-peak near Fermi energy (E$_F$) for $x$-band
and boarding the total $x$- and $z$- subbands as U increased. An interesting
feature of this two-band model is seen in moving the peak from occupied $z$%
-bandstowards the Fermi level with increasing correlation strength $U$, and
a pinning of the van-Hove peak from the $x$-band just above Fermi level.
This drastic change of the $x$- and $z$- band-shape maximize the particle-hole
interband susceptibility (inversely proportional to the energy distance
between band peaks) and therefore increase the fluctuation contribution to
FLEX-energy. The value of the average inverse mass enhancement factor $%
Z=\left[ 1-\partial \Sigma \left( \epsilon \right) /\partial \epsilon \right]
_{\epsilon =0}^{-1}$ is equal to 0.45 and 0.83 for $x$ and $z$ orbitals for
U=4. Spectral function $A({\bf k},\epsilon )=-1/\pi TrImG{\bf (k,\epsilon )}$
for three different ${\bf k}$-directions in two-dimensional Brillouin zone
is shown in Fig.\ref{akfl4}. One could see the renormalized dispersion of
two bands: $x$-band from approximately -5 eV at X-point to 5 eV at M-point
and $z$-band from approximately -8 eV at X-point to 3 eV at M-point. Similar
to all High-T$_c$ models\cite{Licht-QP} there is an extended van-Hove
singularity in $x$-bands just at the Fermi energy near X-point.

The results of self-consistent DMFT calculations for $U=4$ and $8$ are
presents in Fig.\ref{dosdm}. In this case we use only first seven Matsubara
frequencies in the Eq.\ref{Bom} and the constant B-matrix for the rest
frequencies. One can clearly see some differences to the corresponding FLEX
results, which is related to a sharpness of the DOS near E$_F$ and more
pronounced three peak structure of partial $x$-band DOS for U=8. We plot
also the DOS corresponding to the atomic Green function for U=8 with the
four poles near to Fermi energy out of eight poles in the paramagnetic
two-orbital atom. Corresponding spectral function for the same directions in
Brillouin zone is presents in the Fig.\ref{akdm8} for U=8. 
There is a sharp quasiparticle dispersion near the Fermi level and broad
incoherent background above $E_F$ at the energy of about 7 eV near M-point.
Extended van-Hove singularity at X-point become more pronounced. We plot
also the momentum distribution function $n({\bf k})$ in (1,1) direction in
the Fig.\ref{nkdm8}. From the quasiparticle dispersion of the $x$-band along 
$\Gamma -M$ direction (Fig.\ref{akdm8}), we expect the Fermi surface
crossing almost exactly at the half-way between this two points. The momentum
distribution function (Fig.\ref{nkdm8}) just confirm this situations and
shows that the Fermi step (our simulation temperature T=0.06t) is smaller than
one and agree with the calculated value of the mass renormalization factor
Z=0.43 for $x$-band.

Finally the HIA-solution for this two-band model for $U=8$ is shown in Fig.%
\ref{dosh1}. In this case we have the dielectric DOS with narrow
atomic-like resonances. It is interesting to mention that the structure of
the atomic Green function in the DMFT approximation (Fig.\ref{dosdm}) quite
close to the HIA-solution shifted approximately by 2 eV down. We would like
to mention that the application of HIA-scheme is reasonable only for U$>>$W
and not for U=W as in this case.

\section{LDA++ calculations for real systems}

The self-consistent LDA++ calculation for real systems pose a serious
computational problem. One need to operate with the susceptibility matrix,
which is of dimension $N_d^2\times N_d^2$ and depends on the Matsubara
frequencies. For a illustrative purpose we have calculated the electronic
structure of classical Mott-Hubbard insulator NiO and mixed valence
4f-compound TmSe in HIA scheme for LDA++. We use non-selfconsistent HIA
approximation with the simplest constrain for only diagonal $\varepsilon _f$
in Eq.\ref{H-at} to have $n_d$ for NiO or $n_f$ electrons for TmSe from
self-consistent paramagnetic LDA-calculations. The Coulomb matrix is
expressed via effective Slater integrals: 
\[
U_{m_1m_2m_1^{\prime }m_2^{\prime 
}}=\sum_ka_k(m_1,m_1^{\prime },m_2,m_2^{\prime})F^k 
\]
where $0\leq k\leq 2l$ and

\[
a_k(m_1,m_1^{\prime },m_2,m_2^{\prime})=\frac{4\pi }{%
2k+1}\sum_{q=-k}^k\langle lm_1\mid Y_{kq}\mid lm_1^{\prime }\rangle \langle
lm_2\mid Y_{kq}^{*}\mid lm_2^{\prime }\rangle 
\]

We used the following effective Slater parameters, which define screened
Coulomb interaction in d-shell for NiO: $F^0=8.0$ eV, $F^2=8.2$ eV, $F^4=5.2$
eV, and in f-shell for TmSe: $F^0=5.7$ eV, $F^2=9.1$ eV, $F^4=5.7$ eV, $F^6=4.7$
eV, (see e.g. \cite{Anisimov-LDAU,Marel-U,RE-U}). We start from the
non-magnetic LDA calculations in the LMTO nearly-orthogonal representation\cite
{OKA-exact} for experimental crystal structures of NiO and TmSe. The minimal
basis set of $s,p,d$-orbitals for NiO and $s,p,d,f$- orbitals for TmSe
corresponds to 18x18 and 32x32 matrix of the LDA Hamiltonian $h({\bf k)}$.
The occupation number for correlated electrons are 8.4 electrons in the
d-shell of Ni and 12.6 electrons if the 4f-shell of Tm. Using the
corresponding atomic self energy for Ni-atom and Tm-atom the total DOS for
NiO and TmSe have been calculated from Eq.\ref{Hub-1}. In Fig.\ref{nio} we
compare the paramagnetic LDA results with HIA LDA++ scheme. It is well known
that paramagnetic LDA calculations can not produce the insulating gap in
nickel oxide: the Fermi level located in the middle of the half-filled $e_g$
bands\cite{Anisimov-rev}. In the HIA approximation to LDA++ approach there
is a gap (or pseudogap in the Fig\ref{nio} due to temperature broadening)
 of the order of 3.5 eV even in this ``nonmagnetic'' state. This gap
and the satellites at -5 and -8 eV are related to the structure of atomic
Green function shown in the lower panel of the Fig.\ref{nio}.

In Fig.\ref{tmse} we compare the calculated DOS for TmSe with experimental XPS
spectrum \cite{Campagna-TmSe}. The HIA approximation in this case well
reproduces the ladder-type photoemission spectrum, comes mainly from
combinations of two multiplets structure of Tm$^{2+}(f^{13})$ and Tm$%
^{3+}(f^{12}) $. This example demonstrate how the LDA++ scheme can combine
the many-body atomic physics with band structure methods. Normal LDA-band
structure for rare-earth systems 
corresponds to the narrow f-peak at the Fermi level and could not
describe the experimental XPS-spectrum which has the ``f''-resonances over
wide energy range of the order of 12 eV. At the same time, HIA is not
adequate to describe correctly the fine features of the electron structure
near the Fermi level. It is known \cite{Wachter-TmSe} that TmSe is really
the narrow-gap semiconductor. According to the most developed model approach
to mixed-valence semiconductors \cite{IK-TmSe,VKT} the appearance of this
energy gap is caused by both hybridization and exciton effects due to the
Coulomb attraction of 5d conduction electron and 4f hole. This effect cannot
be described in HIA approximation. Nevertheless we believe that the
description of the electronic structure of f compounds including
mixed-valence ones on the large energy scales is important itself and in this
sense the results presented here demonstrate the usefulness of LDA++
approach for the description of real strongly correlated systems.

\section{{\bf \ } SUMMARY}

We have formulate a general LDA++ scheme which takes into
account dynamical electron fluctuations in the case of correlated d- or
f-states. The most accurate approach is the DMFT-band structure method,
while more simple FLEX and HIA scheme can be as well useful for
investigation of correlation effects in real systems, in the
cases of rather weak and rather strong interaction respectively.
 In principle one could
combine the idea of the bath-Green function in DMFT-scheme with the simple
expression for the self-energy in the FLEX approximation. In this case $%
\widehat{\Sigma }=\widehat{\Sigma }^{FLEX}[{\cal G}_0]$ and we expected
effectively reduce the effects of vertex-corrections in the FLEX scheme.

Here we compare the LDA++ approach with more simple LDA+U one. First of all,
to describe Mott insulators in LDA+U approach (as well as in SIC approach)
it is necessary to assume magnetic and (or) orbital ordering \cite
{Anisimov-rev}. In LDA++ it is possible to consider the {\it paramagnetic}
Mott insulators in the framework of {\it ab initio} calculations. Moreover,
it is possible to obtain not only the Mott-Hubbard gap in the electron
spectrum but also satellites and multiplet structure (see e.g. the results for
TmSe and NiO in the previous Section).

The  correlation effects results from the frequency dependence of
the self energy ( the non-Fermi-step form of the distribution function for
quasiparticles, the mass enhancement, the appearance of many-electron
Kondo resonanses etc) can be obtained and investigated in LDA++ approach but
not in LDA+U. Our results for two-band model provide the interesting
examples of such behavior. In particular it is worthwhile to note such
features as the narrowing of the van Hove singularity and its ``pinning'' to
the Fermi level (which is important for the physics of High Tc
superconductors\cite{Licht-QP} and can be described already in multiband
FLEX approximation), the ``three-peak'' structure of the spectrum in the
vicinity of Mott insulators (``Kondo resonance'' and mid-gap states which
can be described in DMFT approach).

 We hope that the approximations described here may be useful
for the {\it ab initio} calculations of electron structure of a great
variety of strongly correlated electron systems including doped Mott
insulators, rare-earth metals and their compounds (in particular
mixed-valence ones), high-temperature superconductors and many others.

\section{Acknowledgements}

Part of this work has been carried out during the visit of one of the author
(MIK) to the Max-Planck-Institute of Solid State (Stuttgart),
Max-Planck-Institute of Physics of Complex Systems (Dresden), and
Forschungszentrum J\"ulich. The authors are grateful to Max-Planck Society
and Forschungszentrum J\"ulich and benefit a lot from discussions with O.K.
Andersen, V.I. Anisimov, A.O. Anokhin, P. Fulde, O. Gunnarsson, L. Hedin, P.
Horsh, and G. Kotliar.

\section{Appendix A}

Here we present the proof of the equivalence of a ''cavity method'' and
impurity problem in DMFT for multiband case. We start with the expression
for the Green function matrix on the zero site in the cavity method (see \cite
{Kotliar-rev}, Eq.(35))

\[
\left[ {\cal G}_0(i\omega )\right] ^{-1}=i\omega +\mu -h_{at}-R, 
\]
where 
\begin{equation}
R=\sum\limits_{ij}t_{0i}G_{ij}^{(0)}t_{j0}  \label{A1}
\end{equation}

with $h_{at}$ being the one-electron part of the intra-atomic Hamiltonian.
All this functions are the matrix in $(m,m^{\prime })$ indices as well as
diagonal matrix in spin-space. Note that $G_{ij}^{(0)}$ is the Green
function between sites $i$ and $j$ on the lattice with the site zero being
eliminated

\begin{equation}
G_{ij}^{(0)}=G_{ij}-G_{i0}G_{00}^{-1}G_{0j}  \label{A2}
\end{equation}

Using the Fourier expansion of all the quantities over the Brillouin zone
and substituting Eq.\ref{A2} to Eq.\ref{A1} one has

\[
R=M-LG_{00}^{-1}L^T 
\]
where 
\[
L=\sum\limits_it_{0i}G_{i0}=\sum\limits_{{\bf k}}t({\bf k})G({\bf k}) 
\]
and 
\[
M=\sum\limits_{ij}t_{0i}G_{ij}t_{j0}=\sum\limits_{{\bf k}}t({\bf k})G({\bf k}%
)t({\bf k}) 
\]

At the same time (see \cite{Kotliar-rev})

\[
G({\bf k})=\left[ \Lambda -t({\bf k})\right] ^{-1} 
\]

\begin{equation}
\Lambda =i\omega +\mu -h_{at}-\Sigma (i\omega )  \label{A3}
\end{equation}

Taking into account that $\sum\limits_{{\bf k}}t({\bf k})=0$ one obtains

\[
L=\sum\limits_{{\bf k}}\left[ t({\bf k})-\Lambda +\Lambda \right] \left[
\Lambda -t({\bf k})\right] ^{-1}=-1+\Lambda G_{00} 
\]
and 
\[
L^T=-1+G_{00}\Lambda 
\]

One can obtain by the similar way the result for M-matrix: $M=\Lambda L^T$

Substituting all this formulas to Eq.\ref{A1} we have

\[
R=(\Lambda +G_{00}^{-1}-\Lambda )L^T=-G_{00}^{-1}+\Lambda 
\]

using Eq.\ref{A3}, we can finally write:

\[
\left[ {\cal G}_0(i\omega )\right] ^{-1}=i\omega +\mu -h_{at}-\Lambda
+G_{00}^{-1}=\Sigma (i\omega )+G_{00}^{-1} 
\]

It means that ${\cal G}_0(i\omega )$ is the Green function of the impurity
problem with the on-site one-electron Hamiltonian $h_{at}+\Sigma (i\omega )$
on each ``non-zero'' site and $h_{at}$ for zero site.

%{\bf Appendix napisan simvolicheski i neliteraturno!!! K tomu zhe u menia
%pod rukoi net obzora Kotliara. Prover pozhaluista vnimatelno, dobav
%neobhodimie poiasnenia i podprav!!!}

\begin{figure}[tbp]
\caption{Energy spectrum for two-band two-site Anderson model in exact
diagonalization and DMFT scheme as well as atomic green function for
correlated cite.}
\label{toy}
\end{figure}

\begin{figure}[tbp]
\caption{Density of states for two band model in the FLEX scheme for
different U-values. Full and dashed lines indicate partial DOS for $x$- and $z$-
orbitals. }
\label{dosfl}
\end{figure}

\begin{figure}[tbp]
\caption{Spectral function (the two band FLEX model, U=4) for three
different directions in the two-dimensional square Brillouin zone: $%
\Gamma=(0,0), X=(0.5,0),$ and $M=(0.5, 0.5)$ in units of $2\pi /a$. }
\label{akfl4}
\end{figure}

\begin{figure}[tbp]
\caption{Density of states for two band model in the DMFT scheme for
different U-values as well as atomic Green function for U=8. Full and dashed
lines indicate partial DOS for $x$- and $z$- orbitals. }
\label{dosdm}
\end{figure}

\begin{figure}[tbp]
\caption{Spectral function (the two band DMFT model U=8) for three different
direction in the two-dimensional square Brillouin zone }
\label{akdm8}
\end{figure}

\begin{figure}[tbp]
\caption{The momentum distribution function n(${\bf k}$) in (1,1) direction
for the two band DMFT model U=8. }
\label{nkdm8}
\end{figure}

\begin{figure}[tbp]
\caption{Density of states for two band model in the HIA scheme for U=8.
Full and dashed lines indicate partial DOS for x and z orbitals. }
\label{dosh1}
\end{figure}

\begin{figure}[tbp]
\caption{Density of states for paramagnetic nickel oxide in the LDA and HIA
approximations as well as Ni-atom Green function. }
\label{nio}
\end{figure}

\begin{figure}[tbp]
\caption{Density of states for TmSe in HIA sheme in comparisson with
experimenral XPS-spectrum \protect\onlinecite{Campagna-TmSe}
 and results of paramagnetic LDA-calculations. }
\label{tmse}
\end{figure}

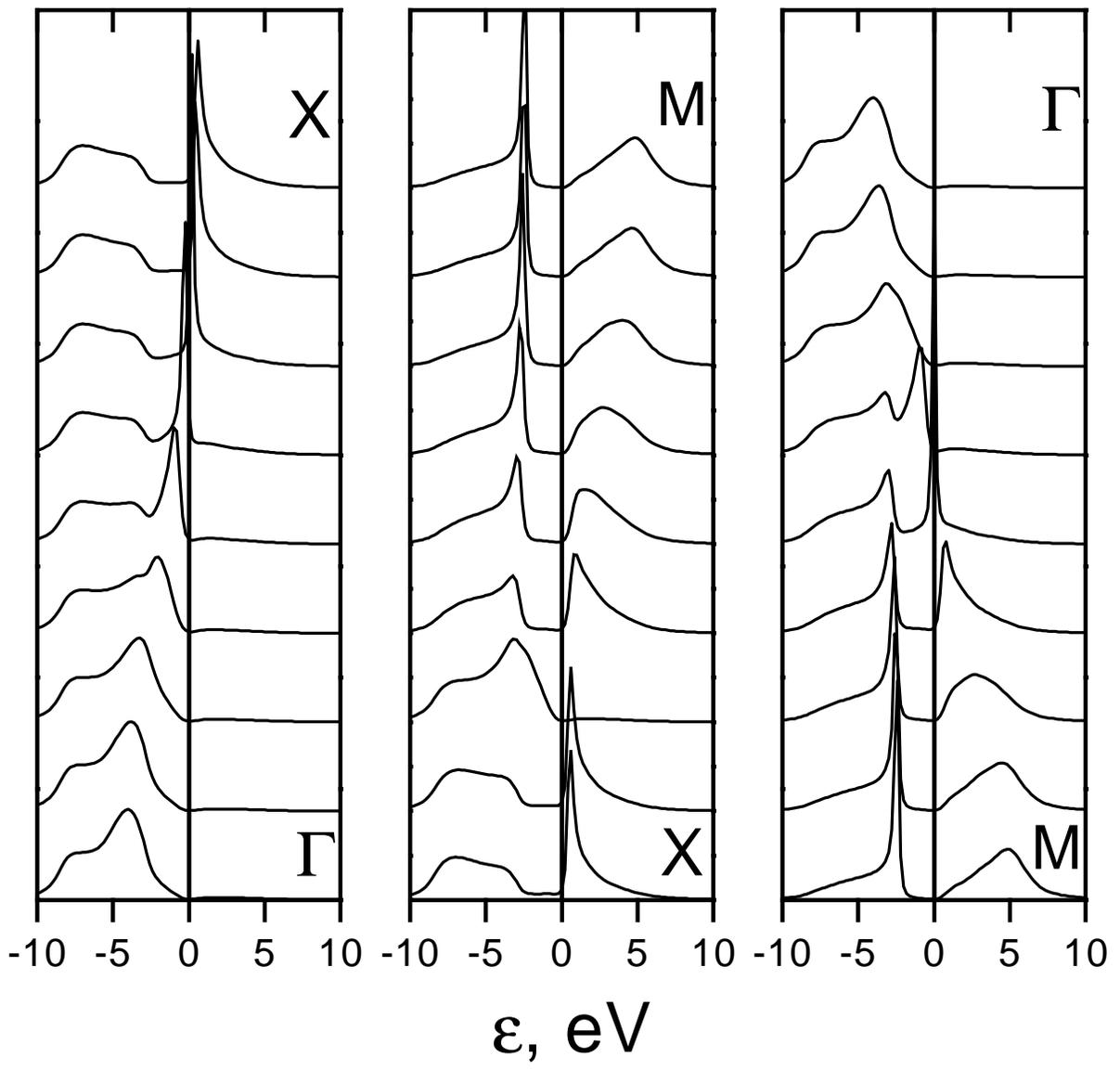
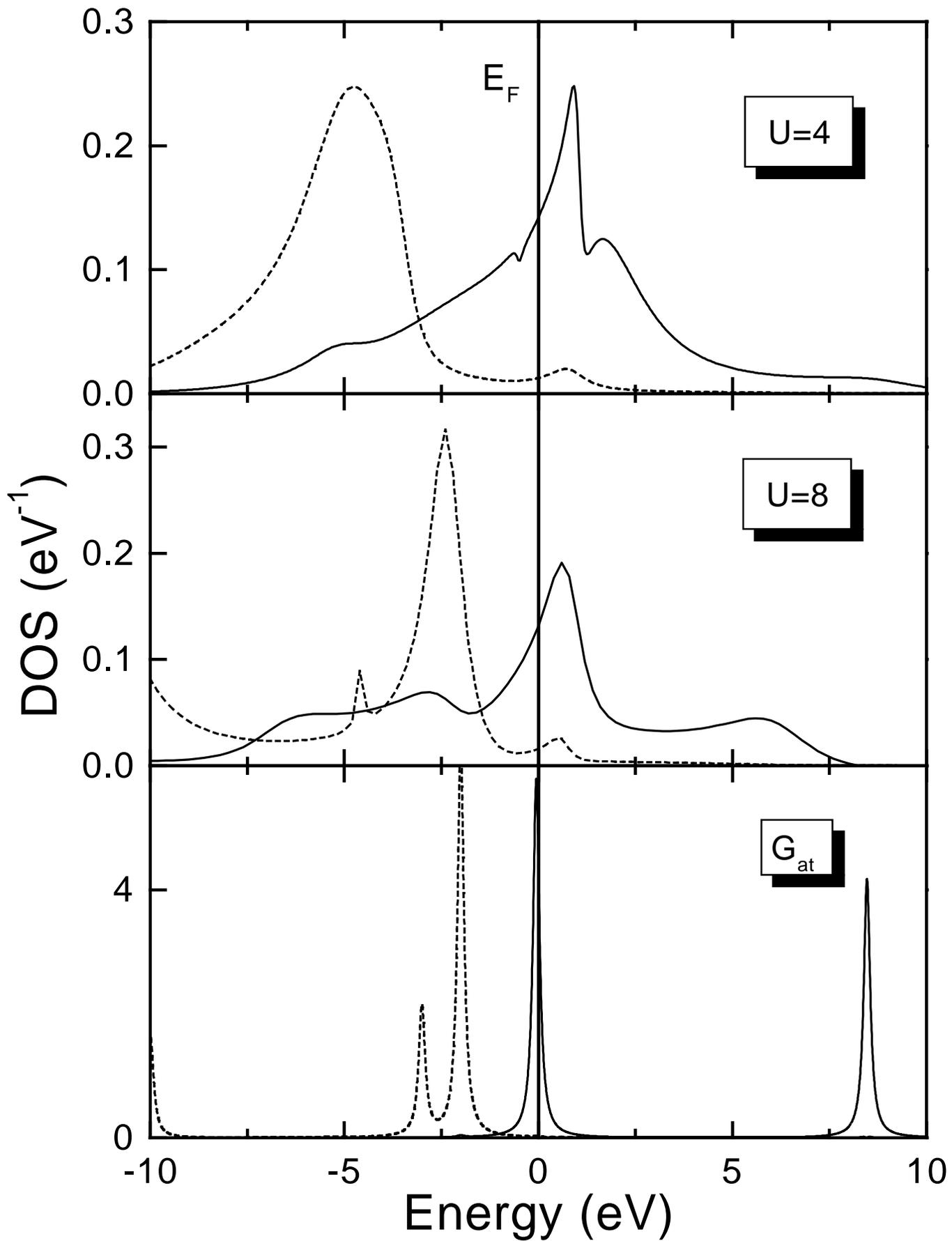
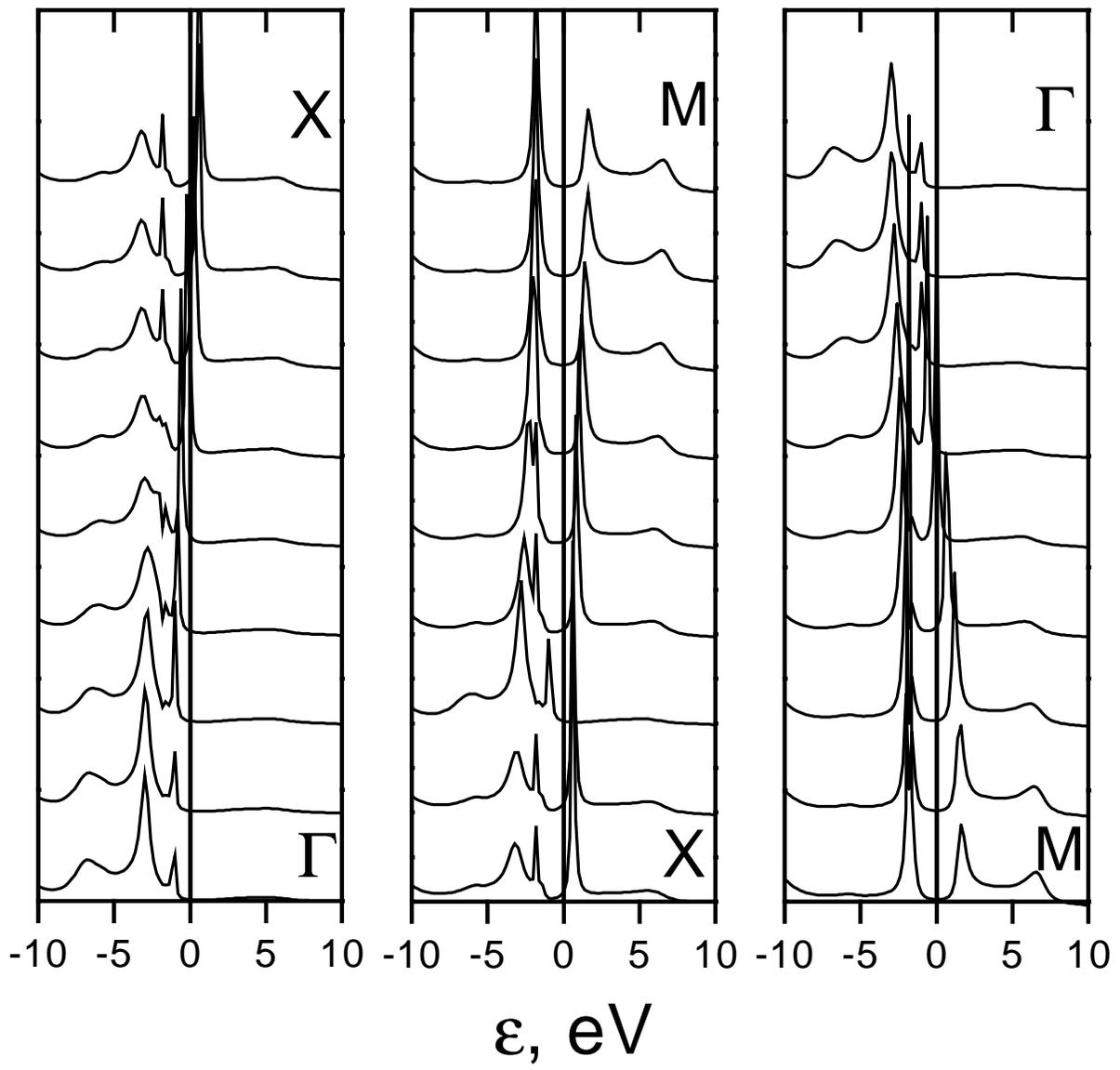
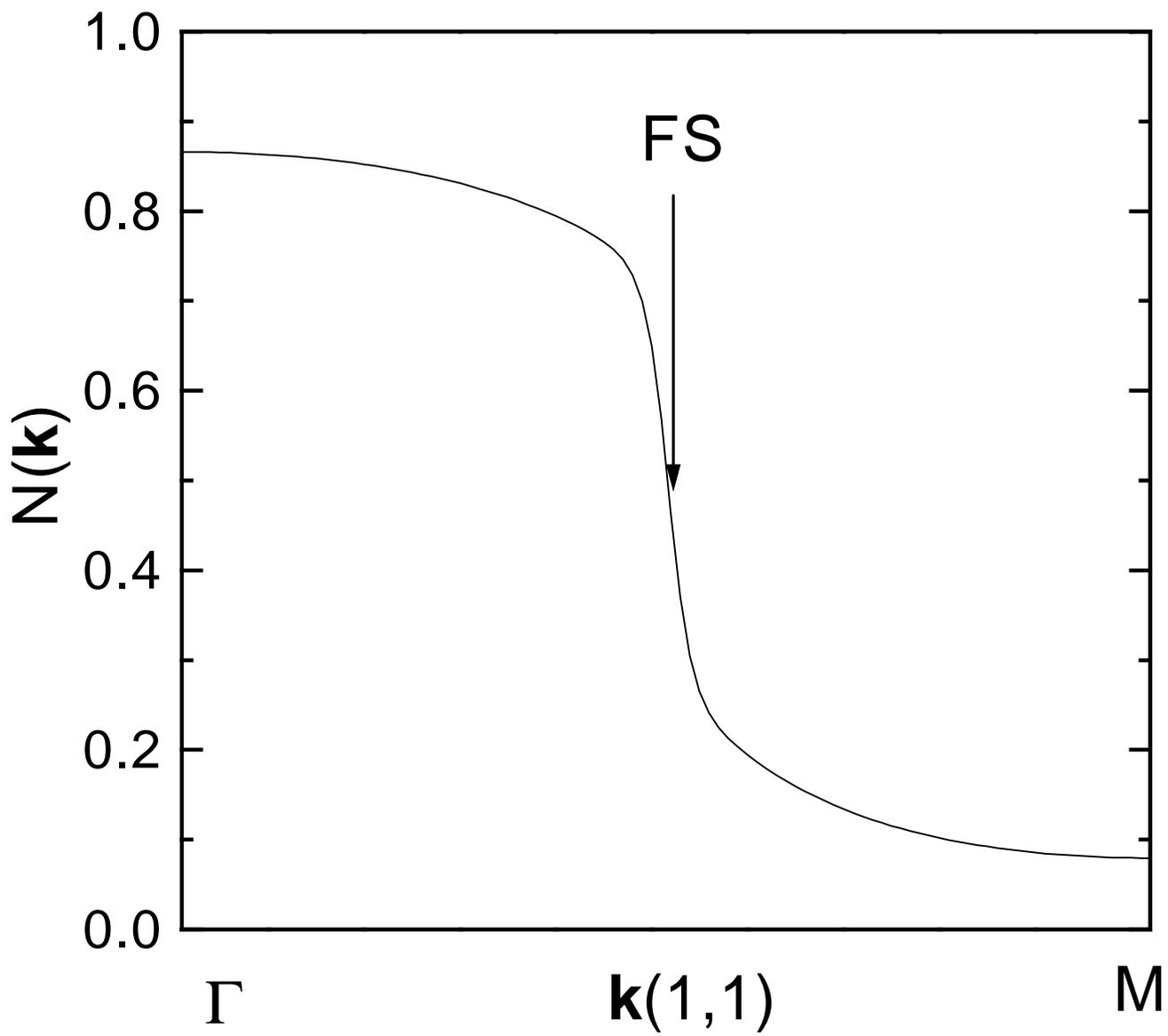
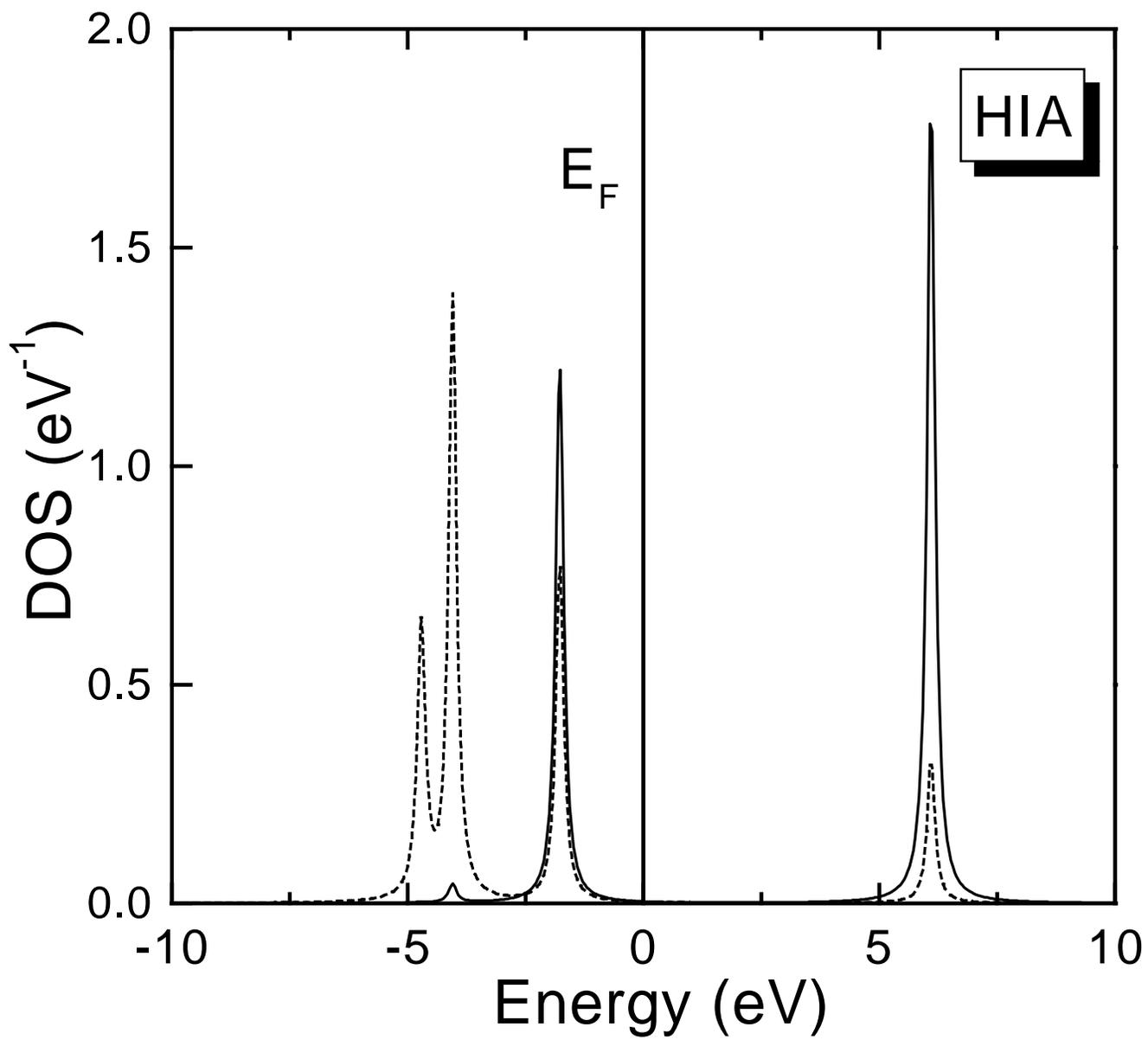
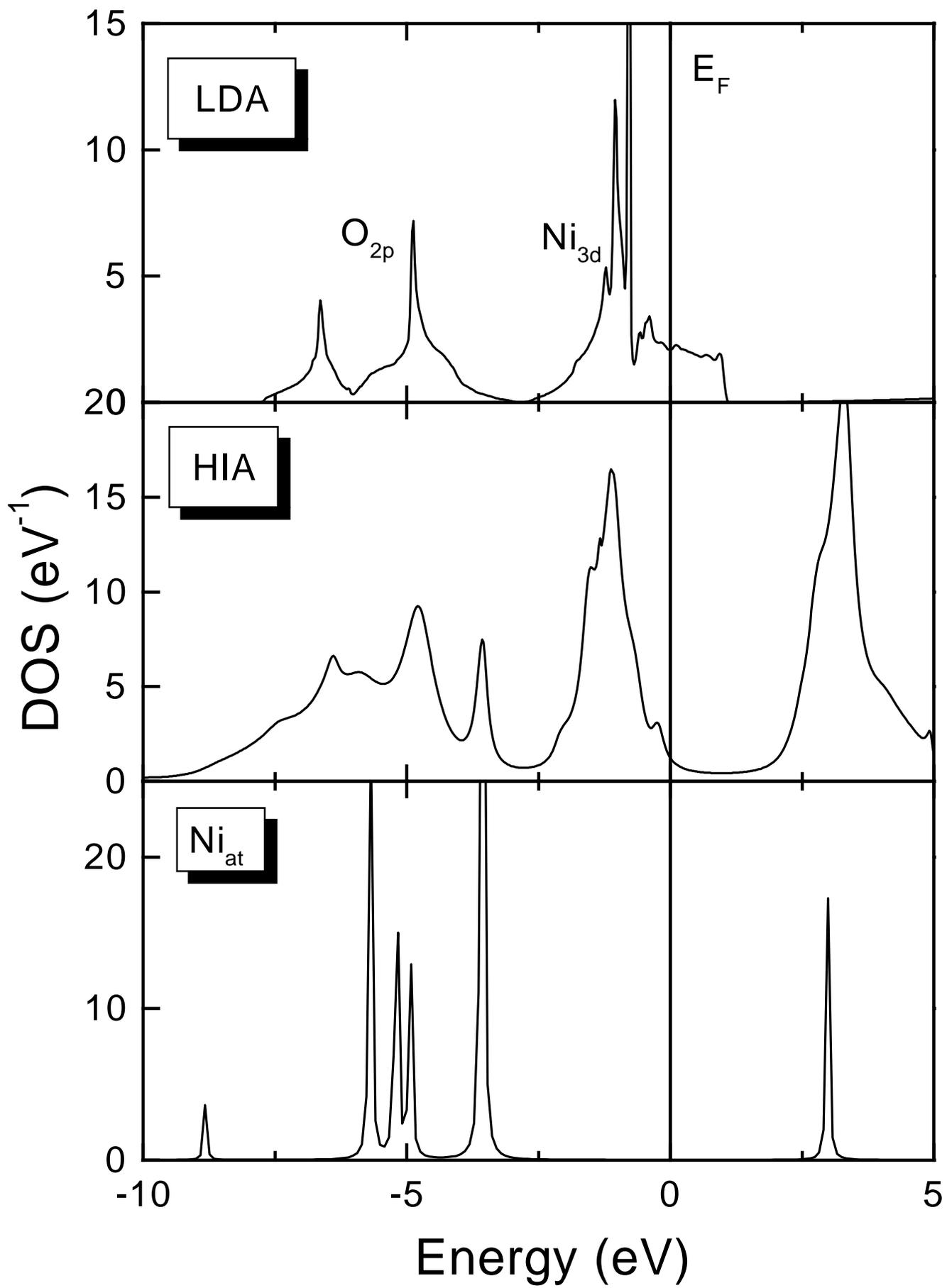
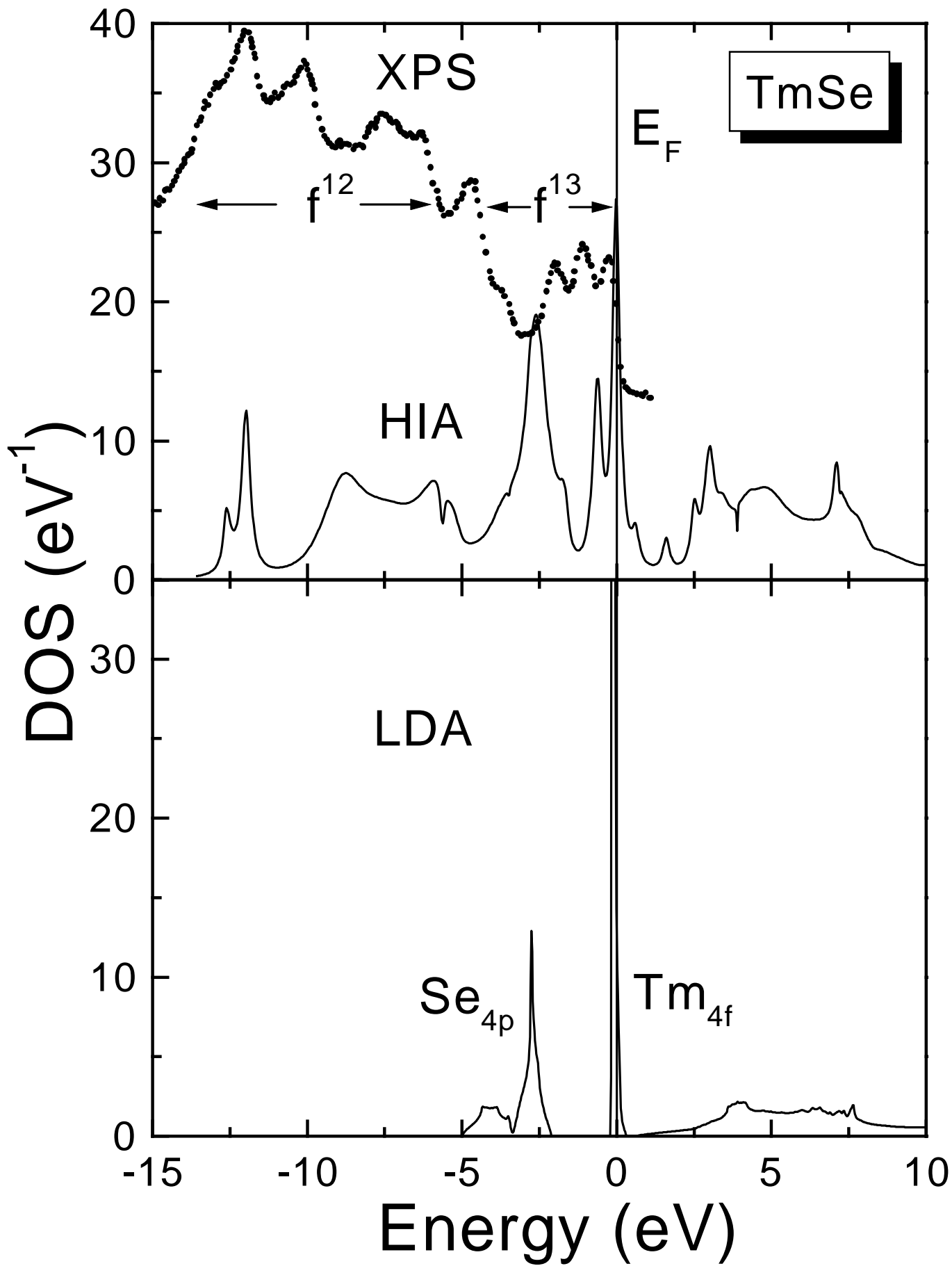
\end{document}